\documentclass{tac}

\usepackage{amssymb}
\usepackage{amsfonts}
\usepackage{latexsym}
\usepackage{epsfig}
\usepackage{diagrams}
\usepackage{verbatim}

\newtheorem{Th}{Theorem}[section]
\newtheorem{theorem}[Th]{Theorem}

\newtheorem{lemma}[Th]{Lemma}

\newtheorem{definition}[Th]{Definition}

\newcommand{\bit}{\begin{itemize}}
\newcommand{\eit}{\end{itemize}\par\noindent}
\newcommand{\ben}{\begin{enumerate}}
\newcommand{\een}{\end{enumerate}\par\noindent}
\newcommand{\beq}{\begin{equation}}
\newcommand{\eeq}{\end{equation}\par\noindent}
\newcommand{\beqa}{\begin{eqnarray*}}
\newcommand{\eeqa}{\end{eqnarray*}\par\noindent}
\newcommand{\beqn}{\begin{eqnarray}}
\newcommand{\eeqn}{\end{eqnarray}\par\noindent}
\newcommand{\ie}{\textit{i.e.}~}

\def\PP{{\rm P}}
\def\HH{{\cal H}}
\def\CC{{\bf C}}
\def\II{{\rm I}}
\newcommand{\dd}{\llcorner}
\newcommand{\sdot}{\bullet}
\newcommand{\ddd}{\lrcorner}
\newcommand{\uu}{\ulcorner}
\newcommand{\uuu}{\urcorner}

\author {Samson Abramsky and Bob Coecke}

\thanks{Rick Blute, Sam Braunstein, Vincent Danos, Martin Hyland and
Prakash
Panangaden provided useful feed-back.}

\address{Oxford University Computing Laboratory,\\ Wolfson Building,
Parks Road, Oxford, OX1 3QD, UK.\\}

\title {Abstract Physical Traces}

\copyrightyear{2004}

\eaddress{abramsky@comlab.ox.ac.uk\CR coecke@comlab.ox.ac.uk}

\begin{document}

\maketitle

\begin{abstract}

We revise our `Physical Traces' paper \cite{AbrCoe1}
in the light of the results in \cite{AbrCoe2}.
The key fact is that the notion of a \emph{strongly compact
closed category} allows abstract notions of adjoint, bipartite
projector and inner product to be defined, and their key properties to be
proved.
In this paper we  improve on the definition of strong compact
closure as compared
to the one presented in \cite{AbrCoe2}. This modification enables an
elegant
characterization of strong compact closure in terms of adjoints and a Yanking axiom, and a
better treatment of
bipartite projectors.
\end{abstract}

\section{Introduction}

In \cite{AbrCoe1} we showed that vector space \em projectors \em
\[
{{\rm P}:V\otimes W\to V\otimes W}
\]
which have a one-dimensional subspace
of
$V\otimes W$ as fixed-points, suffice to implement any linear map, and
also the
categorical trace \cite{JSV} of the category $({\bf
FdVec}_\mathbb{K},\otimes)$
of finite-dimensional vector spaces and linear maps over a base field
$\mathbb{K}$. The interest of this is that projectors of this kind arise
naturally in quantum mechanics (for
$\mathbb{K} = \mathbb{C}$), and play a key role in information protocols such as
\cite{BBC} and \cite{Swap}, and also in measurement-based schemes for quantum computation.  We showed how
both the category $({\bf FdHilb},\otimes)$ of finite-dimensional \em
complex \em Hilbert spaces
and linear maps, and the category $({\bf
Rel},\times)$ of relations with the cartesian product as tensor, can be
physically realized in this sense.

In \cite{AbrCoe2} we showed that such projectors can be defined and their
crucial properties  proved at the abstract level of 
\em strongly compact closed categories\em.   This
categorical structure is a major ingredient of the categorical
axiomatization in \cite{AbrCoe2} of quantum theory
\cite{vN}. It captures quantum entanglement and its behavioral
properties \cite{Coe}. In this paper we will improve
on the definition of strong compact closure, enabling a characterization
in terms of
adjoints - in the linear algebra sense, suitably abstracted - and yanking, without explicit
reference to
compact closure, and enabling a nicer treatment of bipartite projectors,
coherent with
the treatment of arbitrary projectors in \cite{AbrCoe2}.

We are then able to show that the constructions in \cite{AbrCoe1}
for realizing arbitrary morphisms and the trace by projectors also carry over to the
abstract
level, and that these constructions admit an information-flow
interpretation
in the spirit of the one for additive traces
\cite{Abr,AHS}. It is the information flow due to (strong) compact
closure which is crucial  for the abstract formulation, and for  the
proofs of correctness of protocols such as quantum
teleportation \cite{AbrCoe2}.

A concise presentation of (very) basic quantum mechanics which supports
the
developments in this paper can be found in \cite{AbrCoe1,Coe}. However, the
reader with a sufficient categorical background might find the abstract
presentation in \cite{AbrCoe2} more enlightening.

\section{Strongly compact closed categories}

As shown in \cite{KellyLaplaza}, in any monoidal category $\CC$, the
endomorphism monoid $\CC (\II , \II )$ is commutative.
Furthermore any
$s:\II\to\II$ induces a natural transformation
\begin{diagram}
s_A :A & \rTo^{\simeq} & \II \otimes \!A & \rTo^{s \otimes 1_A} & \II
\otimes\! A & \rTo^{\!\!\simeq\ } & A\,.\ \ \ \ \
\end{diagram}
Hence, setting $s \sdot f$ for $f \circ s_A=s_B\circ f$ for $f :
A
\rightarrow B$, we have
\[
(s \sdot g)\circ(r \sdot f)=(s\circ r)\sdot(g\circ f)
\]
for $r:\II\to\II$ and
$g:B\to C$. We call the morphisms $s\in \CC (\II ,
\II )$ \em scalars \em and $s \sdot-$ \em scalar multiplication\em. In
$(\mathbf{FdVec}_{\mathbb{K}},\otimes)$, linear maps
$s:\mathbb{K} \to
\mathbb{K}$ are uniquely determined by the image of $1$, and hence
correspond
biuniquely to elements of $\mathbb{K}$. In $(\mathbf{Rel},\times)$, there
are
just two scalars, corresponding to the Booleans $\mathbb{B}$.

Recall from \cite{KellyLaplaza} that a \em compact closed
category \em is a symmetric
monoidal category ${\bf C}$, in which, when ${\bf C}$ is viewed as a
one-object bicategory,
every one-cell $A$ has a left adjoint $A^*$.
Explicitly this means that for each object $A$ of ${\bf C}$ there exists a
\em dual object \em
$A^*$, a \em unit \em
$\eta_A:{\rm I}\to A^*\otimes A$ and a \em
counit \em $\epsilon_A:A\otimes A^*\to {\rm I}$,
and that the diagrams
\beq\label{ccc1}
\begin{diagram}
A&\rTo^{\simeq\ }&A\otimes{\rm
I}&\rTo^{1_A\otimes\eta_A}&A\otimes(A^*\otimes
A)\\
\dTo^{1_A}&&&&\dTo_{\simeq}\\
A&\lTo_{\simeq\ }&{\rm I}\otimes A&\lTo_{\epsilon_A\otimes
1_A}&(A\otimes A^*)\otimes A
\end{diagram}
\eeq
and
\beq\label{ccc2}
\begin{diagram}
A^*&\rTo^{\simeq\ }&\II \otimes A^*&\rTo^{\eta_A\otimes
1_{A^*}}&(A^*\otimes A)\otimes A^*\\
\dTo^{1_{A^*}}&&&&\dTo_{\simeq}\\
A^*&\lTo_{\simeq\ }&
A^*\otimes\II&\lTo_{1_{A^*}\otimes\epsilon_A}&A^*\otimes
(A\otimes A^*)
\end{diagram}
\eeq
both commute. Alternatively, a compact
closed category may be defined as a $*$-autonomous category \cite{Barr}
with a
self-dual tensor, hence a model of `degenerate' linear logic \cite{Seely}.

For each morphism $f:A\to B$ in a compact closed category we can construct
a \em
dual
\em
$f^*$, a
\em name
\em $\uu f \uuu$ and a \em coname \em $\dd f \ddd$, respectively as
\begin{diagram}
B^*&\rTo^{\simeq}&{\rm I}\otimes B^*&\rTo^{\eta_A\otimes
1_{B^*}}&A^*\otimes
A\otimes B^*\\
\dTo^{f^*}&&&&\dTo_{1_{A^*}\!\otimes f\otimes 1_{B^*}\hspace{-1.3cm}}\\
A^*&\lTo_{\simeq}&A^*\otimes {\rm I}&\lTo_{1_{A^*}\otimes
\epsilon_B}&A^*\otimes
B\otimes B^*
\end{diagram}
\begin{diagram}
A^*\!\!\otimes\! A&\rTo^{1_{A^*}\!\!\otimes\! f}&A^*\!\otimes\! B&&&&&{\rm
I}\\
\uTo^{\eta_A}&\ruTo_{\uu f\uuu}&&&&&\ruTo^{\dd f\ddd}&\uTo_{\epsilon_B} \\
{\rm I}&&&&&A\!\otimes\! B^*&\rTo_{f\!\otimes\! 1_{B^*}}&B\!\otimes\!
B^*&&
\end{diagram}
In particular, the assignment $f\mapsto f^*$ extends $A\mapsto A^*$ into a
contravariant endofunctor with $A\simeq A^{**}$. In any
compact closed category, we have
\[ \CC (A \otimes B^\ast , \II ) \simeq \CC (A, B) \simeq \CC (\II ,
A^\ast \otimes B)\,,
\]
so `elements' of $A \otimes B$ are in biunique correspondence
with names/conames of morphisms $f : A \to B$

Typical examples are $({\bf Rel},\times)$ where $X^*=X$ and where for
$R\subseteq X\times Y$,
\beqa
&&\hspace{-6mm}\uu R\uuu=\{(*,(x,y))\mid xRy,x\in X, y\in Y\}\\
&&\hspace{-6mm}\dd R\ddd=\{((x, y),*)\mid xRy,x\in X, y\in Y\}
\eeqa
and, $({\bf FdVec}_\mathbb{K},\otimes)$ where $V^*$ is the dual vector
space
of linear functionals $v:V\to \mathbb{K}$ and where for $f:V\to W$
with matrix $(m_{ij})$ in bases
$\{e_i^V\}_{i=1}^{i=n}$ and $\{e_j^W\}_{j=1}^{j=m}$ of $V$ and $W$
respectively we have
\beqa
&&\hspace{-6mm}\uu f\uuu:\mathbb{K}\to V^*\otimes
W::1\mapsto\sum_{i,j=1}^{\!i,j=n,m\!}m_{ij}\cdot
\bar{e}_i^V\otimes e_j^W\\ &&\hspace{-6mm}\dd f\ddd:V\otimes
W^*\to\mathbb{K}::e_i^V\otimes\bar{e}_j^W\mapsto m_{ij} .
\eeqa
where $\{\bar{e}_i^{V}\}_{i=1}^{i=n}$ is the base
of $V^*$ satisfying $\bar{e}_i^{V}(e^{V}_j)=\delta_{ij}$, and
similarly for $W$.
Another example is the category $n${\bf Cob} of
$n$-dimensional \em cobordisms \em which is regularly considered in
mathematical
physics, e.g.~\cite{Baez}.

Each compact closed category admits a categorical trace, that is, for
every
morphism
$f:A\otimes C\to B\otimes C$ a trace
${\rm Tr}_{A,B}^C(f):A\to B$ is specified and satisfies certain axioms
\cite{JSV}.  Indeed, we can set
\beq\label{eq:trace}
{\rm Tr}_{A,B}^C(f):=\rho^{-1}_B\circ(1_B\otimes\epsilon_C)\circ (f\circ
1_{C^*})\circ (1_A\otimes (\sigma_{C^*\!,C}\circ\eta_C))\circ\rho_A
\eeq
where $\rho_X:X\simeq X\otimes\II$ and $\sigma_{X,Y}:X\otimes Y\simeq
Y\otimes X$. In $({\bf Rel},\times)$ this yields
\[
x\,{\rm Tr}_{X,Y}^Z(R)y\ \Leftrightarrow\ \exists z\in
Z.(x,z)R(y,z)
\]
for $R\subseteq (X\times Z)\times (Y\times Z)$
while in $({\bf FdVec}_\mathbb{K},\otimes)$ we obtain
\[
{\rm Tr}^U_{V,W}(f):e_i^V\mapsto {\sum}_\alpha m_{i\alpha j\alpha}\, e_j^W
\]
where $(m_{ikjl})$ is the matrix of $f$ in bases $\{e_i^V\!\otimes
e_k^U\}_{ik}$ and
$\{e_j^W\!\otimes e_l^U\}_{jl}$.

\begin{definition}[Strong Compact Closure I]\label{def:sccc1}
\em A \em strongly compact closed category \em is a compact closed
category ${\bf C}$ in
which
$A=A^{**}$ and $(A\otimes B)^*\!\!=A^*\otimes B^*$,  and which comes
together with an
involutive covariant compact closed functor $(\ )_*:{\bf C}\to {\bf C}$
which assigns
each object $A$ to its dual $A^*$.
\end{definition}

So in a strongly compact closed category we have two involutive functors,
namely a contravariant one $(\ )^*:{\bf C}\to {\bf C}$ and a covariant one
$(\ )_*:{\bf
C}\to {\bf C}$ which coincide in their action on objects. Recall that
 $(\ )_*$ being \em compact closed functor \em means that it preserves the
monoidal
structure strictly, and unit and counit i.e.
\beq\label{sccceq}
\uu 1_{A_*}\uuu=(\uu 1_A\uuu)_*\circ u_\II^{-1}\qquad{\rm and}\qquad\dd
1_{A_*}\ddd=u_\II\circ(\dd
1_A\ddd)_*
\eeq
where $u_\II:\II^*\simeq\II$.   This in particular implies that $(\ )_*$
commutes with
$(\ )^*$ since $(\
)^*$ is definable in terms of the monoidal structure, $\eta$ and
$\epsilon$ --- in
\cite{AbrCoe2} we only assumed commutation of $(\ )_*$ and $(\ )^*$
instead of the
stronger requirement of equations (\ref{sccceq}).

For each morphism $f:A\to B$ in a strongly compact closed category we can
define an \em adjoint \em --- as in linear algebra --- as
\[
f^\dagger:=(f_*)^*=(f^*)_*:B\to A\,.
\]
It turns out that we can also define strong compact closure by taking the
adjoint to be a
primitive.

\begin{theorem}[Strong Compact Closure II]\label{def:SCC1}
A strongly compact closed category can be equivalently defined as a
symmetric monoidal category $\CC$ which comes with
\ben
\item a monoidal involutive assignment
$A\mapsto A^*$ on objects,
\item an identity-on-objects, contravariant, strict monoidal, involutive functor
$f\mapsto f^\dagger$, and,
\item for each object $A$ a unit
$\eta_A:\II\to A^*\otimes A$ with $\eta_{A^*}=\sigma_{A^*\!,A}\circ\eta_A$
and such that either the diagram
\beq\label{sccc1}
\begin{diagram}
A&\rTo^{\simeq}&A\otimes{\rm
I}&\rTo^{1_A\otimes\eta_A}&A\otimes(A^*\otimes
A)\\
\dTo^{1_A}&&&&\dTo_{\simeq}\\
A&\lTo_{\simeq}&{\rm I}\otimes
A&\lTo_{(\eta_A^\dagger\circ\sigma_{A,A^*})\otimes 1_A}&(A\otimes
A^*)\otimes A
\end{diagram}
\eeq
or the diagram
\beq\label{sccc2}
\begin{diagram}
A&\rTo^{\!\!\!\!\!\!\simeq\!\!}&{\rm I}\otimes
A&\rTo^{\eta_A\otimes 1_A}&(A^*\!\otimes A)\otimes
A&\rTo^{\simeq}&A^*\!\otimes(A\otimes
A)\\
\dTo^{1_A}&&&&&&\dTo~{1_{A^*}\!\otimes\sigma_{A,A}\!\!\!}\\
A&\lTo_{\!\!\!\!\!\!\simeq\!\!}&{\rm I}\otimes
A&\lTo_{\eta_A^\dagger\otimes 1_A}&(A^*\!\otimes A)\otimes
A&\lTo_{\simeq}&A^*\!\otimes(A\otimes A)
\end{diagram}
\eeq
commutes, where $\sigma_{A,A}:A\otimes A\simeq A\otimes A$ is the twist
map.
\een
\end{theorem}

While diagram (\ref{sccc1}) is the analogue to diagram (\ref{ccc1}) with
$\eta_A^\dagger\circ\sigma_{A,A^*}$ playing the role of the coname,
diagram
(\ref{sccc2}) expresses \em yanking \em with respect to the canonical
trace of
the compact closed structure.  We only need one
commuting diagram as compared to diagrams (\ref{ccc1}) and (\ref{ccc2}) in
the
definition of compact closure and hence in Definition \ref{def:sccc1}
since due
to the strictness assumption (i.e.~$A\mapsto A^*$ being involutive) we
were able
to replace the second diagram by
$\eta_{A^*}=\sigma_{A^*\!,A}\circ\eta_A$.

Returning  to the main issue of this paper, we are now able to
construct a
\em bipartite projector \em (\ie a projector on an object of type
$A\otimes B$)
as
\[
\PP_f:=\uu f\uuu\circ(\uu
f\uuu)^\dagger=\uu f\uuu\circ\dd f_*\ddd:A^*\otimes B\to A^*\otimes B\,,
\]
that is, we have an assignment
\[
\PP_{\_}:\CC(\II,A^*\otimes B)\longrightarrow \CC(A^*\otimes B,A^*\otimes
B)::\Psi\mapsto
\Psi\circ \Psi^\dagger
\]
from bipartite elements to bipartite projectors. Note that the use of $(\ )_*$ is essential in order for $\PP_f$ to be
endomorphic.

We can \em normalize \em these projectors $\PP_f$ by considering
$s_f\sdot
\PP_f$ for
$s_f:=(\dd f_*\ddd\circ\uu f\uuu)^{-1}$ (provided this inverse exists
in $\CC (\II ,
\II )$), yielding
\[
(s_f\sdot \PP_f)\circ(s_f\sdot \PP_f)=s_f\sdot(\uu f\uuu\circ
(s_f\sdot(\dd
f_*\ddd\circ\uu f\uuu))\circ\dd f_*\ddd) =s_f\sdot \PP_f\,,
\]
and also
\[
(s_f\sdot \PP_f)\circ\uu f\uuu=\uu f\uuu\qquad{\rm and}\qquad
\dd f_*\ddd\circ(s_f\sdot \PP_f)=\dd f_*\ddd\,.
\]

Any compact closed category in which $(\ )^{\ast}$ is the identity on
objects is trivially strongly compact closed.
Examples include relations and finite-dimensional real inner-product
spaces, and also the interaction category SProc from \cite{}.

So,
importantly, are finite-dimensional
\emph{complex} Hilbert spaces and linear maps $({\bf FdHilb},\otimes)$. We
take
${\cal H}^*$ to be the \emph{conjugate space}, that is, the Hilbert space with
the same
elements as
${\cal H}$ but with the scalar multiplication and the inner-product in
${\cal H}^*$ defined by
\[
\alpha \sdot_{\HH^*} \phi := \bar{\alpha} \sdot_{\HH} \phi \qquad
\qquad\qquad
\langle \phi \mid \psi \rangle_{\HH^*} := \langle \psi \mid \phi
\rangle_{\HH}\,,
\]
where $\bar{\alpha}$ is the complex conjugate of $\alpha$.
Hence we can still take $\epsilon_\HH$ to be the \em sesquilinear \em
inner-product.

Conversely, an \em abstract notion of inner product \em can be
defined in any strongly compact closed category. Given `elements'
$\psi,\phi:\II\to A$, we define
\[
\langle\psi\mid\phi\rangle:=\psi^\dagger\circ\phi\,\in\CC(\II,\II)\,.
\]
As an example, the inner-product in
$(\mathbf{Rel},\times)$ is, for $x,y\subseteq\{*\}\times X$,
\[
\langle x\mid y\rangle= 1_{\II} \quad{\rm for}\quad x \cap y \neq
\varnothing
\qquad
{\rm and}
\qquad
\langle x\mid y\rangle= 0_{\II} \quad{\rm for}\quad x \cap y = \varnothing
\]
with
$1_{\II}:=\{*\}\times\{*\}\subseteq\{*\}\times\{*\}$ and
$0_{\II}:=\emptyset\subseteq\{*\}\times\{*\}$.
When defining
\em unitarity of an isomorphism \em $U:A\to B$ by $U^{-1}=U^\dagger$ we
can prove
the defining properties both of inner-product space adjoints and
inner-product
space unitarity:
\[
\langle
f^\dagger\!\circ\psi\mid\phi\rangle_B=(f^\dagger\!\circ\psi)^\dagger\!
\circ\phi=
\psi^\dagger\!\circ f\circ\phi=\langle \psi\mid
f\circ\phi\rangle_A\,,
\]
\[
\langle U\circ\psi\mid U\circ\varphi\rangle_B=
\langle U^\dagger\!\circ U\circ\psi\mid \varphi\rangle_A=
\langle \psi\mid \varphi\rangle_A\,,
\]
for $\psi,\varphi:\II\to A$, $\phi:\II\to B$, $f:B\to A$ and $U:A\to B$.
As shown in \cite{AbrCoe2}, an alternative way to define the abstract
inner-product is
\begin{diagram}
\ \II \!& \rTo^{\rho_{\II}} & \II \otimes \II & \rTo^{1_\II \!\otimes\!
u_{\II}} & \II
\otimes
\II^*\! & \rTo^{\phi \!\otimes\! \psi_*} & A \otimes A^*\! &
\rTo^{\epsilon_A} &
\!\II
\end{diagram}
where $u_\II:\II\simeq \II^*$ and $\rho_\II:\II\simeq \II\otimes\II$. Here
the
key  data we use is  the coname $\epsilon_A:A \otimes A^*\to\II$\,,
and also
$(\ )_*$: cf.~also the above examples of both real and complex
inner-product
spaces where $\epsilon_A:=\langle-\mid-\rangle$.
Hence it is fair to say that
\[
{{\rm strong\ compact\ closure}\over{\rm compact\ closure}}
\ \simeq\
{{\rm inner\mbox{\rm -}product\ space}\over{\rm vector\ space}}\,.
\]

Finally, note that abstract bipartite projectors $\PP_f$ have two
components: a `name'-component and a `coname'-component. While in most
algebraic
treatments involving projectors these are taken to be primitive, in our
setting
projectors
 are composite entities, and this decomposition will carry over to their
crucial
properties (see below). We
depict names, conames, and projectors as follows:

\vspace{3.5mm}\noindent{
\begin{minipage}[b]{1\linewidth}
{\epsfig{figure=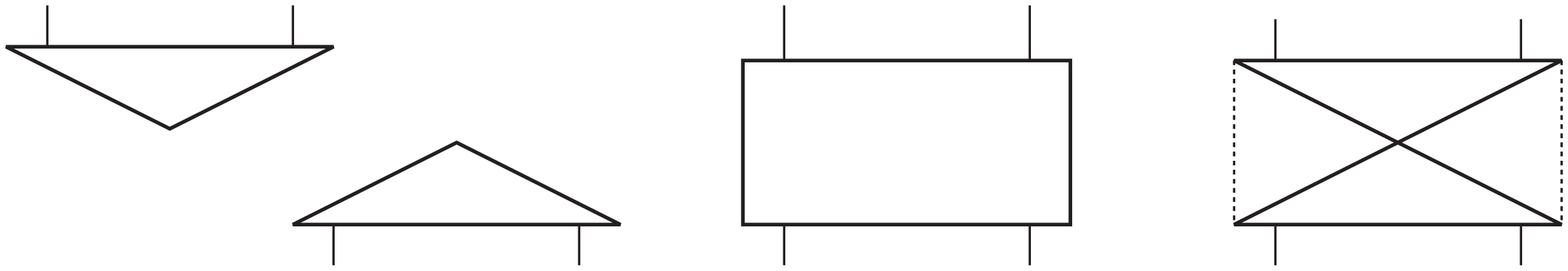,width=380pt}}

\begin{picture}(380,0)
\put(33.5,55.5){$\uu f\uuu$}
\put(102.5,30){$\dd f\ddd$}
\put(216.5,41){\large$\PP_f$}
\put(273,41){\LARGE$:=$}
\put(330.5,53){$\uu f\uuu$}
\put(330.5,30){$\dd f_{\!*\!}\ddd$}
\end{picture}
\end{minipage}}

\vspace{-2.5mm}\noindent
In this representation, diagrams (\ref{ccc1}) and (\ref{sccc2}) can be
expressed as the
respective pictures

\vspace{3.0mm}\noindent{
\begin{minipage}[b]{1\linewidth}
\centering{\epsfig{figure=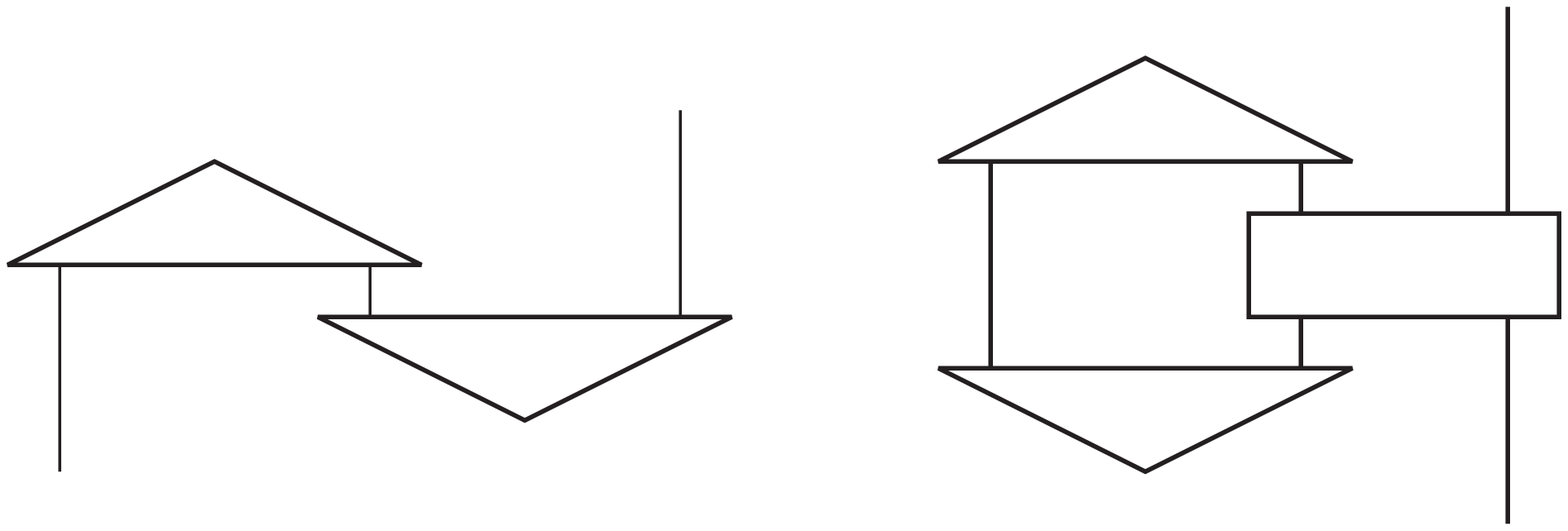,width=300pt}}

\begin{picture}(300,0)
\put(33.5,71.5){$\ \epsilon_A$}
\put(93.5,43.5){$\ \eta_A$}
\put(211,90.5){$\ \eta_A^\dagger$}
\put(212.5,33.5){$\ \eta_A$}
\put(260.5,63.5){$\ \sigma_{A,A}$}
\end{picture}
\end{minipage}}

\vspace{-4mm}\noindent
being equal to the identity.  Below we will express equalities in this
manner.

\section{Information-flow through projectors}

\begin{lemma}[Compositionality - Abramsky and Coecke LiCS`04]
In a compact closed category
\[
\lambda^{-1}_C\circ (\dd f\ddd\otimes 1_C)\circ(1_A\otimes\uu
g\uuu)\circ\rho_A=g\circ f
\]
for $A\rTo^{f}B\rTo^{g}C$, $\rho_A:A\simeq
A\otimes\II$ and
$\lambda_C:C\simeq\II\otimes C$, i.e.,

\vspace{3.0mm}\noindent{
\begin{minipage}[b]{1\linewidth}
\centering{\epsfig{figure=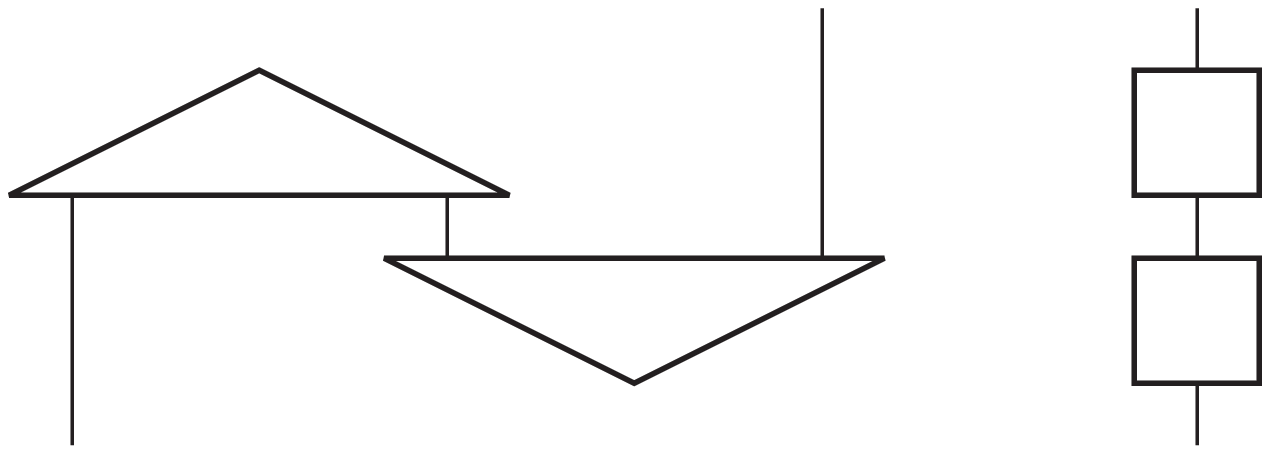,width=200pt}}

\begin{picture}(200,0)
\put(32.5,59.5){$\dd f\ddd$}
\put(93,33.5){$\uu g\uuu$}
\put(153.5,44){\LARGE$=$}
\put(187,62){$g$}
\put(187,31.5){$f$}
\end{picture}
\end{minipage}}

\vspace{-4mm}\noindent
in our graphical representation.
\end{lemma}

Following \cite{AbrCoe2,Coe} we can think of the information
flowing along the grey line in the diagram below, being acted on by
the morphisms which label the coname and the name respectively.

\vspace{0.5mm}\noindent{
\begin{minipage}[b]{1\linewidth}
\centering{\epsfig{figure=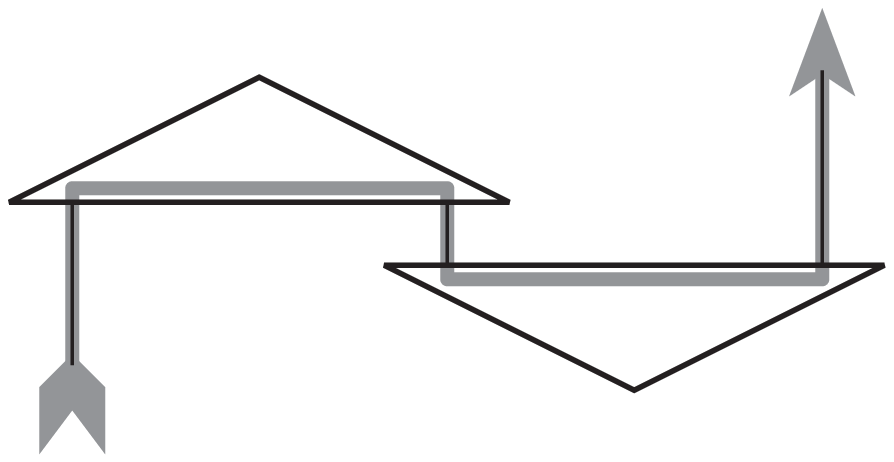,width=140pt}}

\begin{picture}(140,0)
\put(32.5,60.5){$\dd f\ddd$}
\put(91,31){$\uu g\uuu$}
\end{picture}
\end{minipage}}

\vspace{-3.5mm}\noindent
We refer to this as the \em information-flow interpretation of compact
closure\em.  Many variants can also be derived
\cite{AbrCoe2,Coe}. The pictures expressing the non-trivial branches of
diagrams
(\ref{ccc1}) and (\ref{sccc2}) become

\vspace{2.0mm}\noindent{
\begin{minipage}[b]{1\linewidth}
\centering{\epsfig{figure=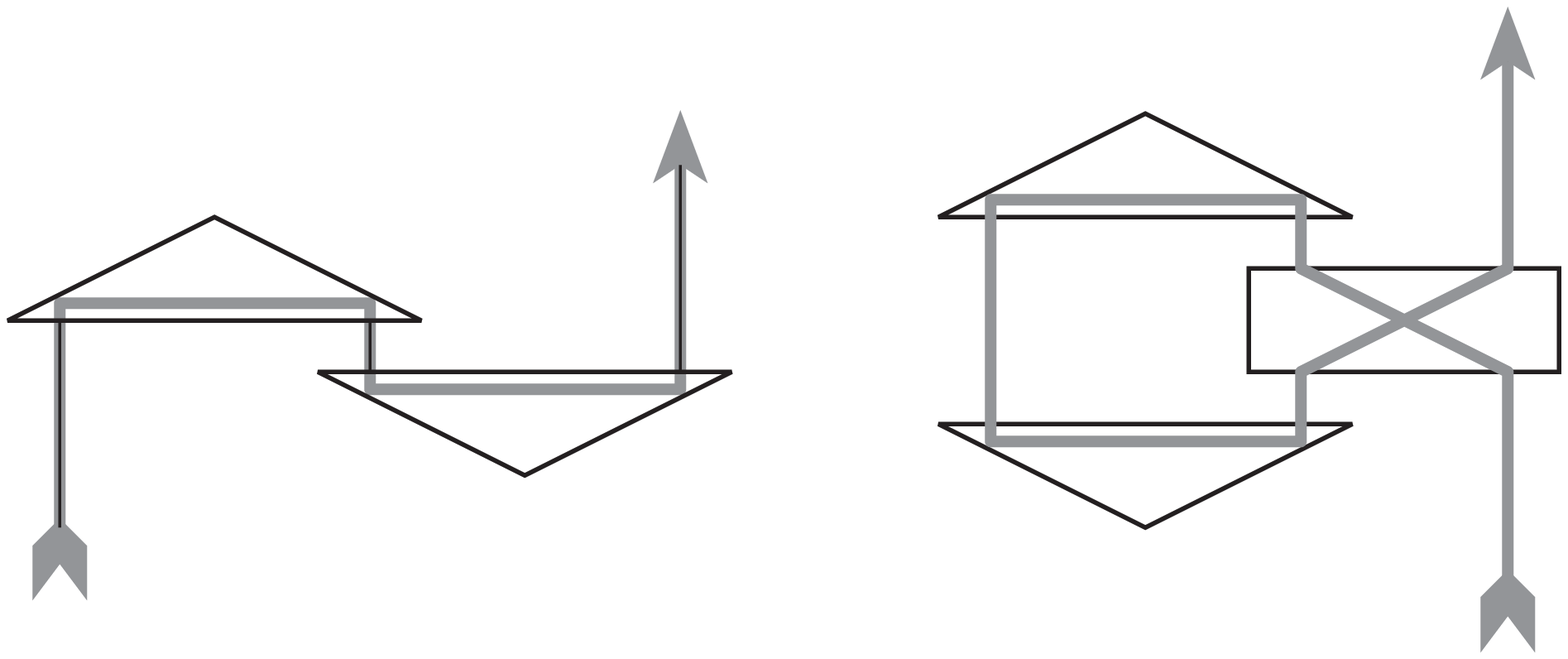,width=300pt}}

\begin{picture}(300,0)
\put(35.5,86.5){$\ \epsilon$}
\put(95.5,56.5){$\ \eta$}
\put(211,106){$\ \eta^\dagger$}
\put(214,46){$\ \eta$}
\put(213.5,470){$\ \eta$}
\put(263.5,81.5){$\ \sigma$}
\end{picture}
\end{minipage}}

\vspace{-5mm}

Lemma 2 of \cite{AbrCoe1}, which states that we can realize any
linear map $g:V\to W$ using only $({\bf
FdHilb},\otimes)$-projectors, follows trivially
by setting
$f:=1_{V}$ while viewing both $\dd 1_{V\!\!}\ddd$
and
$\uu g\uuu$ as being parts of  projectors --- all this is up to a
scalar multiple which depends on the input of
$\PP_g$. Note that by functoriality $1_{V^*}=(1_V)_*$
and hence $\PP_{(1_V)_*}\!\!=\PP_{1_{V^*}}$. As discussed in
\cite{Coe} this feature constitutes the core of
\em logic-gate teleportation\em, which is a fault-tolerant
universal quantum computational primitive \cite{Gottesman}.
Explicitly,

\begin{lemma}
In a strongly compact closed category $\CC$ for $f:A\to B$,
\[
f\otimes(\uu 1_{A^*\!\!}\uuu\circ \dd
\xi\ddd)=s(f,\xi) \sdot\left(\sigma_{A,B}\circ(\PP_{1_{A^*}}\otimes
1_B)\circ(1_A\otimes
\PP_f)\right)
\]
where $s(f,\xi)\in\CC(\II,\II)$ is a scalar,  $\sigma_{A,B}:A\otimes
A^*\!\otimes B\to B\otimes A^*\!\otimes A$ is symmetry, $\xi:A^*\!\to B^*$
is arbitrary, and $s(f,f_*)=1_\II$.
\end{lemma}

\noindent Lemma 1 of \cite{AbrCoe1}, that is, we can realize the $({\bf
FdHilb},\otimes)$-trace by means of projectors
trivially follows from eq.(\ref{eq:trace}), noting that
$\eta=\uu 1\uuu$ and $\epsilon=\dd 1\ddd$ and again viewing
these as parts of  projectors. Explicitly:

\begin{lemma}
In a strongly compact closed category $\CC$ for $f:A\otimes C\to B\otimes
C$,
\[
{\rm Tr}_{A,B}^C(f)\otimes(\uu 1_{C^*}\uuu\circ \dd
\xi\ddd)=s(\xi) \sdot\left((1_A\otimes
\PP_{1_{C^*}})\circ (f\otimes 1_{C^*})\circ(1_B\otimes
\PP_{1_{C^*}})\right)
\]
where $s(\xi)\in\CC(\II,\II)$ is a scalar, $\xi:C\to C$ is
arbitrary, and $s(1_C)=1_\II$.
\end{lemma}

\noindent
Indeed, since $\sigma_{A^*\!,A}\circ\uu 1_A\uuu=\uu
(1_{\!A})^{\!*\!}\uuu=\uu 1_{\!A^*\!}\!\!\uuu$ by functoriality,
eq.(\ref{eq:trace}) is

\vspace{3.5mm}\noindent{
\begin{minipage}[b]{1\linewidth}
\centering{\epsfig{figure=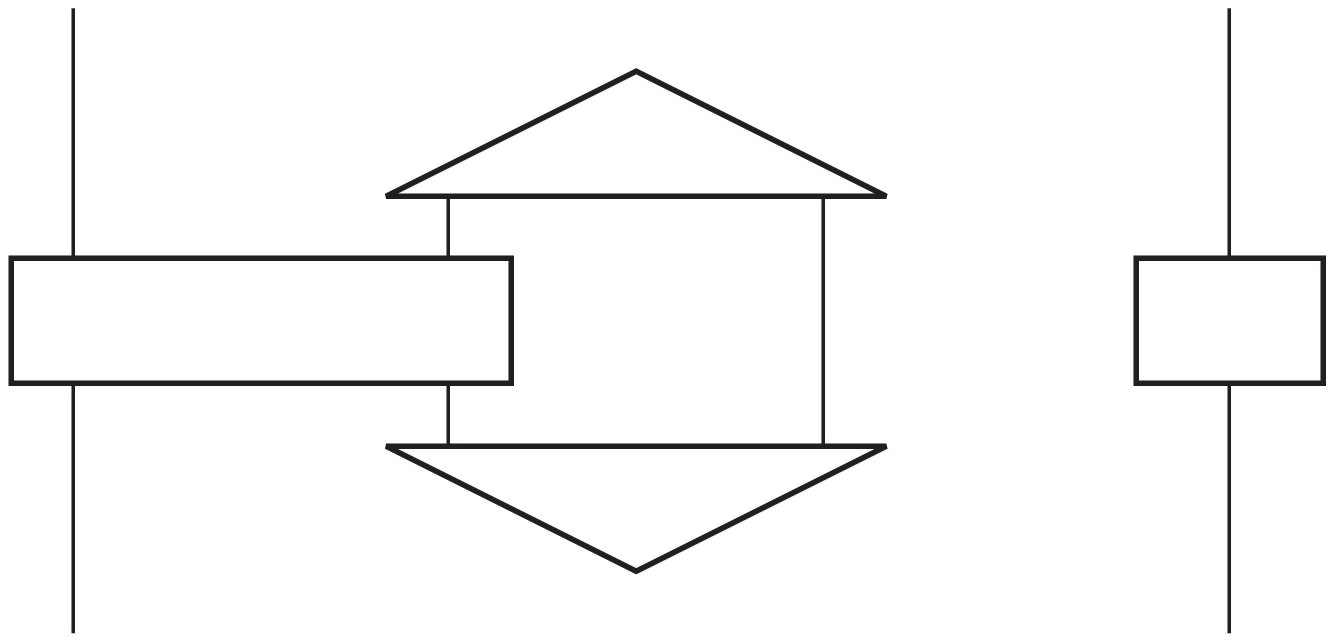,width=210pt}}

\begin{picture}(210,0)
\put(37.5,62){$f$}
\put(92,33.5){$\uu 1\uuu$}
\put(92,90.5){$\dd 1\ddd$}
\put(151.5,59){\LARGE$=$}
\put(182,61.5){${\rm Tr}(f)$}
\end{picture}
\end{minipage}}

\vspace{-3.5mm}\noindent
Interestingly, using the information-flow
interpretation of compact closure,
provided $f$  itself admits an information-flow interpretation, this
construction
admits one too, and can be regarded as a feed-back construction. As an
example,
for $f:=(g_1\otimes g_2)\circ\sigma\circ(f_1\otimes f_2)$, we
have (use naturality of $\sigma$, the definition of (co)name and
compositionality)

\vspace{3.5mm}\noindent{
\begin{minipage}[b]{1\linewidth}
\centering{\epsfig{figure=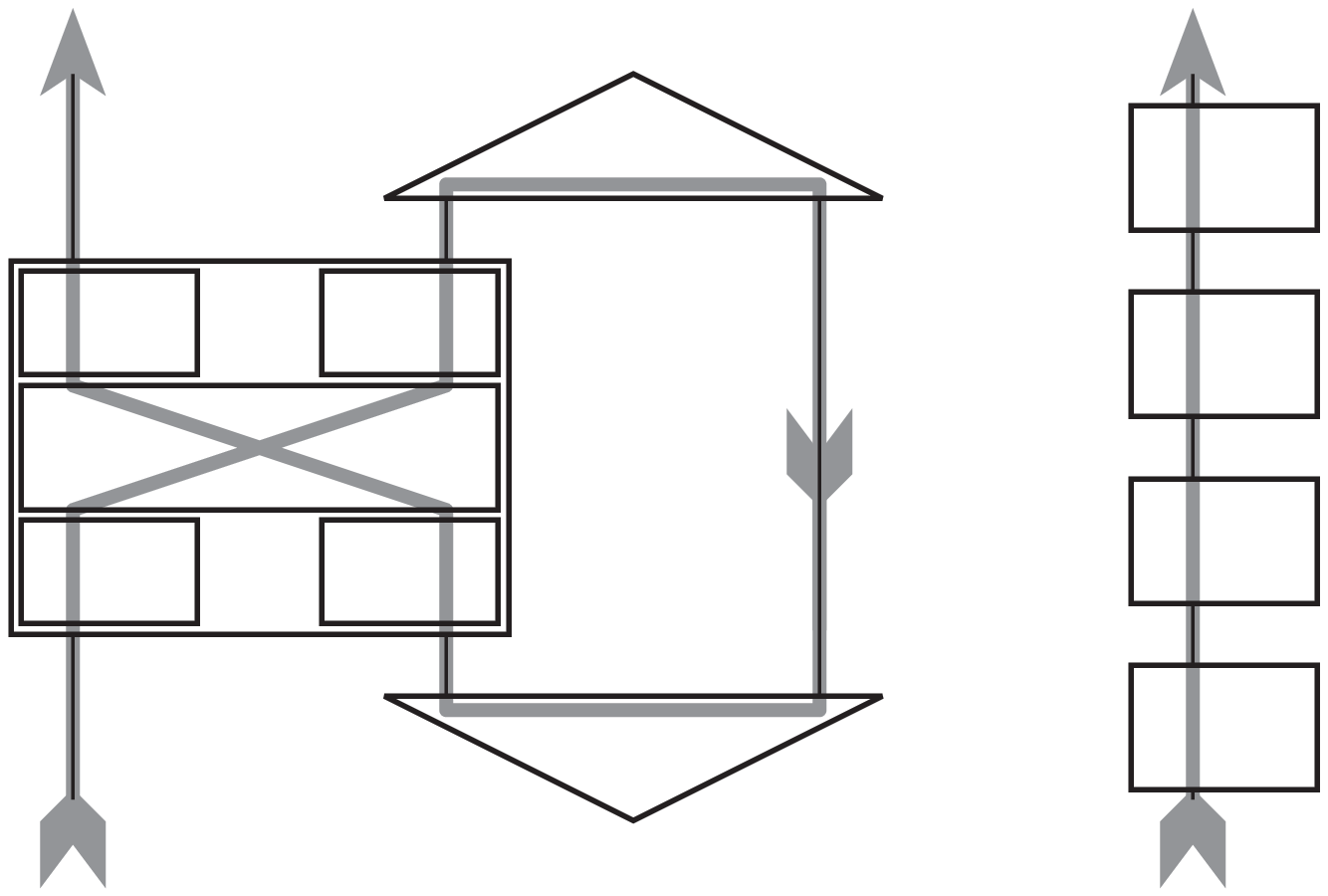,width=210pt}}

\begin{picture}(210,0)
\put(16,62){$f_1$}
\put(57,62){$f_2$}
\put(16,103){$g_1$}
\put(57,103){$g_2$}
\put(38.5,87.7){$\sigma$}
\put(91,31.5){$\uu 1\uuu$}
\put(91,131.5){$\dd 1\ddd$}
\put(151.5,79){\LARGE$=$}
\put(196,37){$f_1$}
\put(196,67){$g_2$}
\put(196,97){$f_2$}
\put(196,127){$g_1$}
\end{picture}
\end{minipage}}

\vspace{-3.5mm}\noindent
When taking $f$ itself to be a projector
$\PP_g=\uu g\uuu\circ\dd g_*\!\ddd$ we have

\vspace{3.5mm}\noindent{
\begin{minipage}[b]{1\linewidth}
\centering{\epsfig{figure=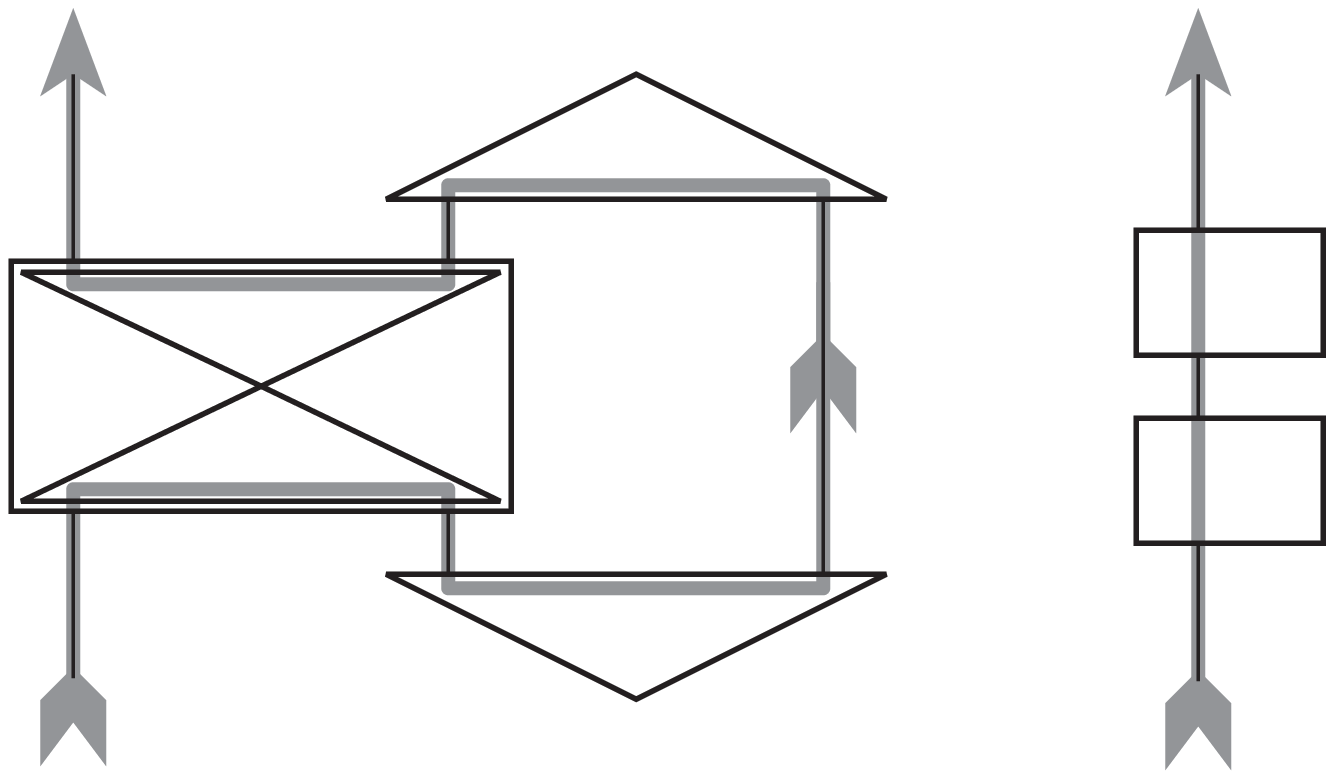,width=210pt}}

\begin{picture}(210,0)
\put(32,81){$\uu f\uuu$}
\put(32,63.5){$\dd f_{\!*\!}\ddd$}
\put(91.5,31.5){$\uu 1\uuu$}
\put(91.5,112.5){$\dd 1\ddd$}
\put(151.5,71){\LARGE$=$}
\put(196,58){$f_*$}
\put(196,88){$f^*$}
\end{picture}
\end{minipage}}

\vspace{-3.5mm}\noindent
using $\sigma\circ\uu f\uuu=\uu f^{\!*\!}\uuu$, naturality of
$\sigma$ and compositionality.
Note that the information-flow in the loop is in this case
`forward' as compared to `backward' in the previous example.
For $f$ of type $A\otimes (C_1\otimes \ldots \otimes
C_n)\to B\otimes (C_1\otimes \ldots \otimes C_n)$ we can have
multiple looping:

\vspace{3.5mm}\noindent{
\begin{minipage}[b]{1\linewidth}
\centering{\epsfig{figure=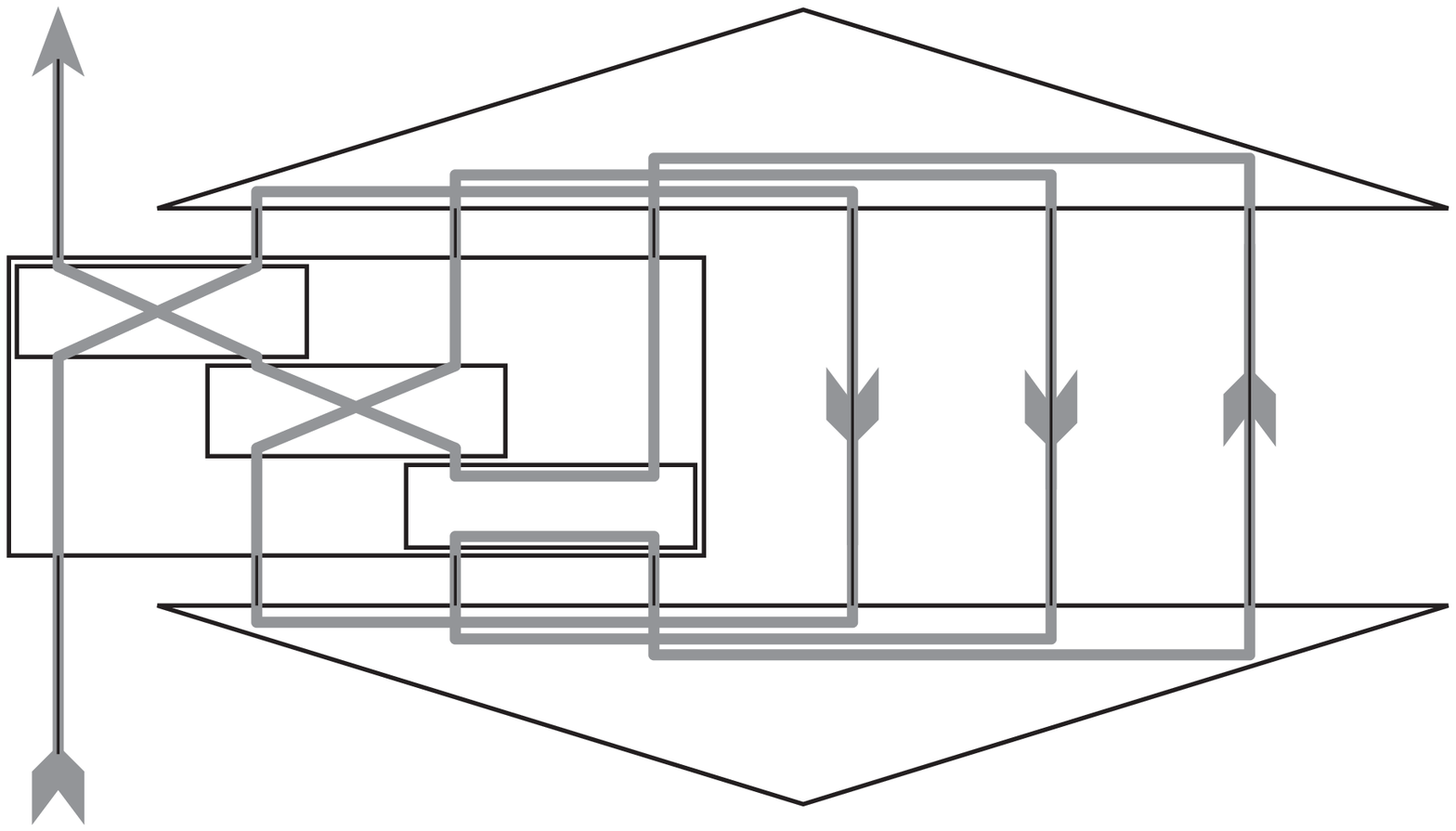,width=290pt}}

\begin{picture}(290,0)
\put(107.8,74.3){$\PP$}
\put(28.3,120.0){$\sigma$}
\put(68.3,100.0){$\sigma$}
\put(152,31.5){$\uu 1\uuu$}
\put(152,157.5){$\dd 1\ddd$}
\end{picture}
\end{minipage}}

\noindent Note the resemblance between this behavior and that of \em
additive
traces \em \cite{Abr,AHS} such as the one on $({\bf Rel},+)$ namely
\[
x\,{\rm Tr}_{X,Y}^Z(R)y\Leftrightarrow \exists z_1,
\ldots, z_n\in Z.xRz_1R \ldots R z_n Ry
\]
for $R\subseteq X+Z\times Y+Z$. In this case we can
think of a particle traveling trough a network where the elements $x\in X$
are
the possible states
of the particle. The morphisms $R\subseteq X\times Y$
are processes that impose a (non-deterministic) change of state $x\in X$
to
$y\in R(x)$, emptyness of $R(x)$ corresponding to undefinedness.
The sum
$X+Y$ is the disjoint union of state sets and
$R+S$ represents parallel composition of processes.  The trace ${\rm
Tr}_{X,Y}^Z(R)$
is \em feedback\em, that is, entering in a state
$x\in X$ the particle will either halt, exit at $y\in Y$ or, exit at
$z_1\in Z$ in which case it is fed back into
$R$ at the
$Z$ entrance, and so on, until it halts or exits at $y\in Y$.

For a more conceptual view of the matter, note that the examples
illustrated above all live in the \emph{free compact closed category}
generated by a suitable category in the sense of \cite{KellyLaplaza}.
Indeed our
diagrams, which are essentially `proof nets for compact closed logic'
\cite{AD}, give a presentation of this free category. Of course, these
diagrams will then have representations in any compact closed
category. For a detailed discussion of free contsructions for traced and strongly compact closed categories, see the forthcoming paper \cite{Abr05}.

\section{$({\bf FRel},\times,{\rm Tr})$ from $({\bf FdHilb},\otimes,{\rm
Tr})$}

In \cite{AbrCoe1} \S3.3 we provided a lax functorial passage from the
category $({\bf
FdHilb},\otimes,{\rm Tr})$ to the category of finite sets and relations
$({\bf FRel},\times,{\rm Tr})$.  This passage involved  choosing  a
base for
each Hilbert space.
When restricting the morphisms of
${\bf FdHilb}$ to those for which the matrices in the chosen bases are
$\mathbb{R}^+$-valued we obtain a true functor.

The results in \cite{AbrCoe2}, together with the ideas developed in this
paper, provide a better understanding of this passage.  In any monoidal
category, $\CC(\II,\II)$ is an abelian monoid \cite{KellyLaplaza}
(Prop.~6.1).
If $\CC$ has a zero object $0$ and biproducts $\II\oplus...\oplus\II$
we obtain an abelian semiring with zero $0_\II:\II\to\II$ and sum
$-+-:\nabla_\II\circ (-\oplus-)\circ\Delta_\II:\II\to\II$. When in such a
category every object is isomorphic to
one of the form $\II\oplus \cdots \oplus\II$ (finitary), as is the case
for
both $({\bf FdHilb},\otimes)$ and $({\bf FRel},\times)$, then this
category
is equivalent (as a monoidal category) to the category of
$\CC(\II,\II)$-valued
matrices with the usual matrix operations.  Note that this equivalence
involves choosing a basis isomorphism for each object. For $({\bf
FdHilb},\otimes)$ we have
$\CC(\II,\II)\simeq\mathbb{C}$ and for $({\bf FRel},\times)$ we have
$\CC(\II,\II)\simeq\mathbb{B}$, the semiring of booleans. Such a category
of matrices is
trivially strongly compact closed for
$(\bigoplus_{i=1}^{i=n}\II)^*:=\bigoplus_{i=1}^{i=n}\II$,
\[
\eta:=(\delta_{i,j})_{i,j}:\II\to
\left(\bigoplus_{i=1}^{i=n}\II\right)\!\otimes\!
\Biggl(\bigoplus_{j=1}^{j=n}\II\Biggr)
\]
(using distributivity and
$\II\otimes\II\simeq\II$), and
\[
\epsilon:\left(\bigoplus_{i=1}^{i=n}\II\right)\otimes
\left(\bigoplus_{i=1}^{i=n}\II\right)\to\II::
(\psi,\phi)\mapsto \phi^T\circ \psi
\]
where $\phi^T$ denotes the transpose of
$\psi$.
In the case of $({\bf
FRel},\times)$, this yields the strong compact closed structure
described above. If
the abelian semiring $\CC(\II,\II)$ also admits a non-trivial involution
$(\
)_*$, an alternative  compact closed structure arises by defining
$\epsilon::
(\psi,\phi)\mapsto (\phi^T)_*\circ \psi$, where $(\
)_*$ is applied pointwise. The corresponding strong compact closed structure involves defining the adjoint of a matrix $M$ to be $M^T_*$, \ie the involution is applied componentwise to the transpose of $M$.
In this way we obtain (up to categorical equivalence) the strong
compact closed structure on $({\bf FdHilb},\otimes)$ described above,
taking
$(\ )_*$ to be complex conjugation.

Now we can relate trace preserving and (strongly) compact closed functors to (involution preserving) semiring homomorphisms.
Any such homomorphism $h : R \to S$ lifts to a functor on the
categories of matrices. Moreover, such a functor preserves compact
closure (and strong compact closure if $h$ preserves the given
involution), and hence also the trace.
Clearly there is no semiring embedding
$\xi:\mathbb{B}\rightarrow\mathbb{C}$
since $\xi(1+1)\not=\xi(1)+\xi(1)$. Conversely, for
$\xi:\mathbb{C}\rightarrow\mathbb{B}$ neither $\xi(-1)\mapsto 0$ nor
$\xi(-1)\mapsto 1$ provide a true homomorphism. But setting $\xi(c)=1$ for
$c\not=0$
we have $\xi(x+y)\leq \xi(x)+\xi(y)$ and $\xi(x\cdot y)=\xi(x)\cdot\xi(y)$
which
lifts to a \emph{lax} functor --- {\bf FRel} is order-enriched, so this makes sense.
Restricting from $\mathbb{C}$ to
$\mathbb{R}^+$ we obtain a true homomorphism, and hence a compact
closed functor.

\refs

\bibitem [Abramsky 1996]{Abr}
Abramsky, S. (1996) {\em Retracing some paths in process algebra}.
Proceedings of
the Seventh International Conference on Concurrency Theory, LNCS
{\bf 1119}, 1--17.
\bibitem[Abramsky 2005]{abr05}
Abramsky, S. (2005) Abstract Scalars, Loops, and Free Traced and Strongly Compact Closed Categories. To appear.
\bibitem [Abramsky and Coecke CTCS`02]{AbrCoe1}
Abramsky, S. and Coecke, B. (2003) {\em Physical traces:
quantum vs.~classical information  processing}.
Electronic notes on Theoretical Computer  Science {\bf 69}
(CTCS`02 issue). \texttt{arXiv:cs/0207057}
\bibitem [Abramsky and Coecke LiCS`04]{AbrCoe2}
Abramsky, S. and Coecke, B. (2004)
{\em A categorical semantics of quantum protocols}.
Proceedings of the 19th Annual IEEE Symposium
on Logic in Computer Science (LiCS`04), IEEE Computer
Science Press.
(extended version at \texttt{arXiv: quant-ph/0402130})
\bibitem [Abramsky and Coecke (nd)]{AC2}
Abramsky, S. and Coecke, B. (2004)
{\em Abstract quantum mechanics}.
In preparation. (major improvements and additions as compared to
\cite{AbrCoe2})
\bibitem [Abramsky and Duncan 2004]{AD}
Abramsky, S. and Duncan, R.~W. (2004)
\emph{Categorical quantum logic}.
To appear in the Proceedings of the Second International Workshop on
Quantum Programming Languages (QPL 04).
\bibitem [Abramsky, Haghverdi and Scott 2002]{AHS}
Abramsky, S., Haghverdi, E.  and Scott, P.~J. (2002) {\em Geometry of
interaction and
linear combinatory algebras}. Mathematical Structures in Computer Science
{\bf 12},
625--665.
\bibitem [Baez 2004]{Baez}
Baez, J. (2004) {\em Quantum quandaries: a category-theoretic
perspective}.
Structural Foundations of Quantum Gravity, eds. S. French, D. Rickles and
J.
Sahatsi, Oxford University Press. \texttt{arXiv:quant-ph/0404040}
\bibitem [Barr 1979]{Barr}
Barr, M. (1979) {\em $*$-Autonomous Categories}.
Lecture Notes in Mathematics {\bf 752}, Springer-Verlag.
\bibitem [quantum teleportation 1993]{BBC}
Bennet, C.~H., Brassard, C., Cr\'epeau, C., Jozsa, R., Peres, A.  and
Wooters, W.~K. (1993) {\em Teleporting an unknown quantum state via dual
classical
and Einstein-Podolsky-Rosen channels}.  Physical Review Letters {\bf 70},
1895--1899.
\bibitem [Coecke 2003]{Coe}
Coecke, B. (2003)
{\em The logic of entanglement. An invitation}. Research Report
PRG-RR-03-12 Oxford
University Computing Laboratory.
\texttt{web.comlab.ox.ac.uk/oucl/ publications/tr/rr-03-12.html} (short
version at
\texttt{arXiv:quant-ph/0402014})
\bibitem [Gottesman and Chuang 1999]{Gottesman}
Gottesman, D. and Chuang, I.~L. (1999) {\em Quantum teleportation is a
universal
computational primitive}.  Nature {\bf 402}, 390--393.
\texttt{arXiv: quant-ph/9908010}
\bibitem [Joyal, Street and Verity 1996]{JSV}
Joyal, A., Street, R.  and Verity, D. (1996) {\em Traced monoidal
categories}.
Proceedings of the Cambridge Philosophical Society {\bf 119}, 447--468.
\bibitem [Kelly and Laplaza 1980]{KellyLaplaza}
Kelly, G.~M. and Laplaza, M.~L. (1980) {\em Coherence for compact closed
categories}.
Journal of Pure and Applied Algebra {\bf 19}, 193--213.
\bibitem [Seely 1998]{Seely}
Seely, R.~A.~G. (1998) \em Linear logic, {$*$}-autonomous categories and
cofree algebras\em. Categories in Computer Science and Logic, Contemporary
Mathematics {\bf 92}, 371--382.
\bibitem [von Neumann 1932]{vN}
von Neumann, J. (1932) {\it Mathematische Grundlagen der
Quantenmechanik}. Springer-Verlag. English translation (1955):
{\it Mathematical Foundations of Quantum Mechanics}.
Princeton University Press.
\bibitem [entanglement swapping 1993]{Swap}
\.{Z}ukowski, M., Zeilinger, A., Horne, M.~A. and Ekert, A.~K. (1993)
{\em `Event-ready-detectors' Bell experiment via entanglement swapping}.
Physical Review Letters {\bf 71}, 4287--4290.

\endrefs

\end{document}